\documentclass[review]{elsarticle}

\usepackage{amsmath, amssymb}

\usepackage{times}
\usepackage{bm}
\usepackage{natbib}

\usepackage[plain,noend]{algorithm2e}
\usepackage{graphicx,psfrag,epsf}
\usepackage{color} 
\usepackage{enumerate}
\usepackage {textcomp}
\usepackage{qtree}

\newcommand{\bw}{0}

\newdefinition{definition}{Definition}
\newtheorem{lemma}{Lemma}
\newtheorem{theorem}{Theorem}

\journal{Journal of Statistical Planning and Inference}
\begin{document}

\begin{frontmatter}
\title{Equitable $(d,m)$-edge designs}

\author{Jean-Marc~F\'edou\corref{cor1}}
\ead{fedou@unice.fr}
\author{Maria-Jo\~ao Rendas\corref{cor2}}
\ead{rendas@i3s.unice.fr}
\address{Laboratoire I3S - UMR7271 - UNS CNRS
2000, route des Lucioles 06900 Sophia Antipolis - France}

\cortext[cor1]{Principal corresponding author}
\cortext[cor2]{Corresponding author}

\begin{abstract}
The paper addresses design of experiments for  classifying the input factors of a multi-variate function into negligible, linear and other (non-linear/interaction) factors.  We give constructive procedures for completing the definition of the clustered designs proposed in \cite{morris:91}, that become defined for arbitrary number of input factors and desired clusters' multiplicity. Our work is based on a representation of subgraphs of the hyper-cube by polynomials that allows the formal verification of the designs' properties. Ability to generate these designs in a systematic manner opens new perspectives for the characterisation of the behaviour of the function's derivatives over the input space that may offer increased discrimination.
\end{abstract}
\end{frontmatter}


\section{Introduction}
\subsection{Sensitivity analysis}

 In sensitivity analysis, one wishes to characterise the dependency of an unknown function $f: {\cal A} \subset \mathbb{R}^d \rightarrow \mathbb{R}$ on each of its $d$ input factors. In general, we know nothing about the function $f(\cdot)$, but  can evaluate it at chosen locations $\xi\in \mathbb{R}^d$. Interest is on partitioning the factors of $f(\cdot)$ into those that have no impact on the function value (class ${\cal C}_0$), that have a linear effect (class  ${\cal C}_1$) or that are non-linear or have interactions with other input factors (class ${\cal C}_2$). Often, fast screening is done in the context of factor fixing  (as noted in \cite{Saltelli:06}), with the goal of restricting subsequent analysis of $f(\cdot)$ to the smaller set  ${\cal C}_2$. This is the context we address.

Several methods have been proposed for sensitivity analysis, ranging from local  to global methods, in particular, variance based methods such as the use of Monte-Carlo methods for the computation of Sobol indices (\cite{Sobol:93}), the Fourier Amplitude Sentivity Test (FAST) method (\cite{Cukier:73}, \cite{Cukier:78}), or the Morris elementary effect method (\cite{morris:91}).
Morris method for preliminary sensitivity analysis is one of the most commonly used, due to its robustness and computational efficiency. The method has not only been applied to a variety of different fields (see \cite{Saltelli:06} for a review), but has also received the attention of several researchers who proposed modifications and improvements:  \cite{cropp:02}--\cite{campolongo:99} propose  an extension enabling study of two-factor interaction terms, 
\cite{campo:07} chooses the design used to evaluate the elementary effects  amongst a large number of random trajectories, such that a dispersion index is optimised, 
 \cite{pujol:07}  replaces designs aligned with the input space directions by randomly oriented simplexes, and  \cite{bouk:11} proposes a sequential version of Morris test,  so that computational effort is concentrated in class ${\cal C}_2$ factors.
We focus on Morris' original method, that we outline below. Our contribution  concerns  the designs used in Morris method, and can be combined with most modifications of the original method published in the literature.
\subsection{Morris preliminary sensitivity analysis designs}
Morris method implements statistical tests  over a set of elementary differences along each principal direction $i$, $d_{i}(\xi)$, computed at a set of  points $\{\xi_n\}_{n=1}^r$ of the input domain:
\begin{equation}
d_{i}(\xi)= \frac{1}{\Delta} \left[ f(\xi+\Delta e_i) - f(\xi) \right], \ \ \xi \in {\cal A},\ i =1,\ldots,d.
\end{equation}
Above, $e_i$ is the vector  with components $e_{i_j} = \delta_{ij}, j=1,\ldots, d$. Let  $(\mu_i,\sigma^2_i)$ be  empirical estimates of the mean $\overline{d}_i$ and variance $s_i^2$ of  $d_i$:
 \begin{align}
\mu_i &= \frac{1}{r} \sum_{n=1}^r d_i(\xi^i_n) \simeq \mbox{E}_{\nu_i}\left[ d_i\right]=\overline{d}_i,\qquad i=1,\ldots, d\enspace \label{eq:mu}\\
\sigma_i ^2&= \frac{1}{r-1} \sum_{n=1}^r \left( d_i(\xi^i_n) - \mu_i  \right)^2\simeq \mbox{Var}_{\nu_i}\left[ d_i \right]=s_i^2\enspace .\label{eq:sigma2}
 \end{align}
Input factors are classified as  ({\em i}) negligible, ({\em ii}) linear, or  ({\em iii}) non-linear/interaction   if  ({\em i}) their mean and variance are both close to zero,  ({\em ii}) the mean is non-zero, but variance is small, or ({\em iii})  variance is large.  A revised version of Morris method (\cite{campo:07}) uses instead $\mu_i^\star$, the sample average of $|d_i(\xi)|$, improving the robustness for derivatives of alternating sign.

\begin{figure}
\vspace*{-.6cm}
\begin{center}
\if1\bw
{
\includegraphics[width=.45\textwidth]{Morris1bw.pdf}\hspace{\fill}
}\fi
\if0\bw
{
\includegraphics[width=.45\textwidth]{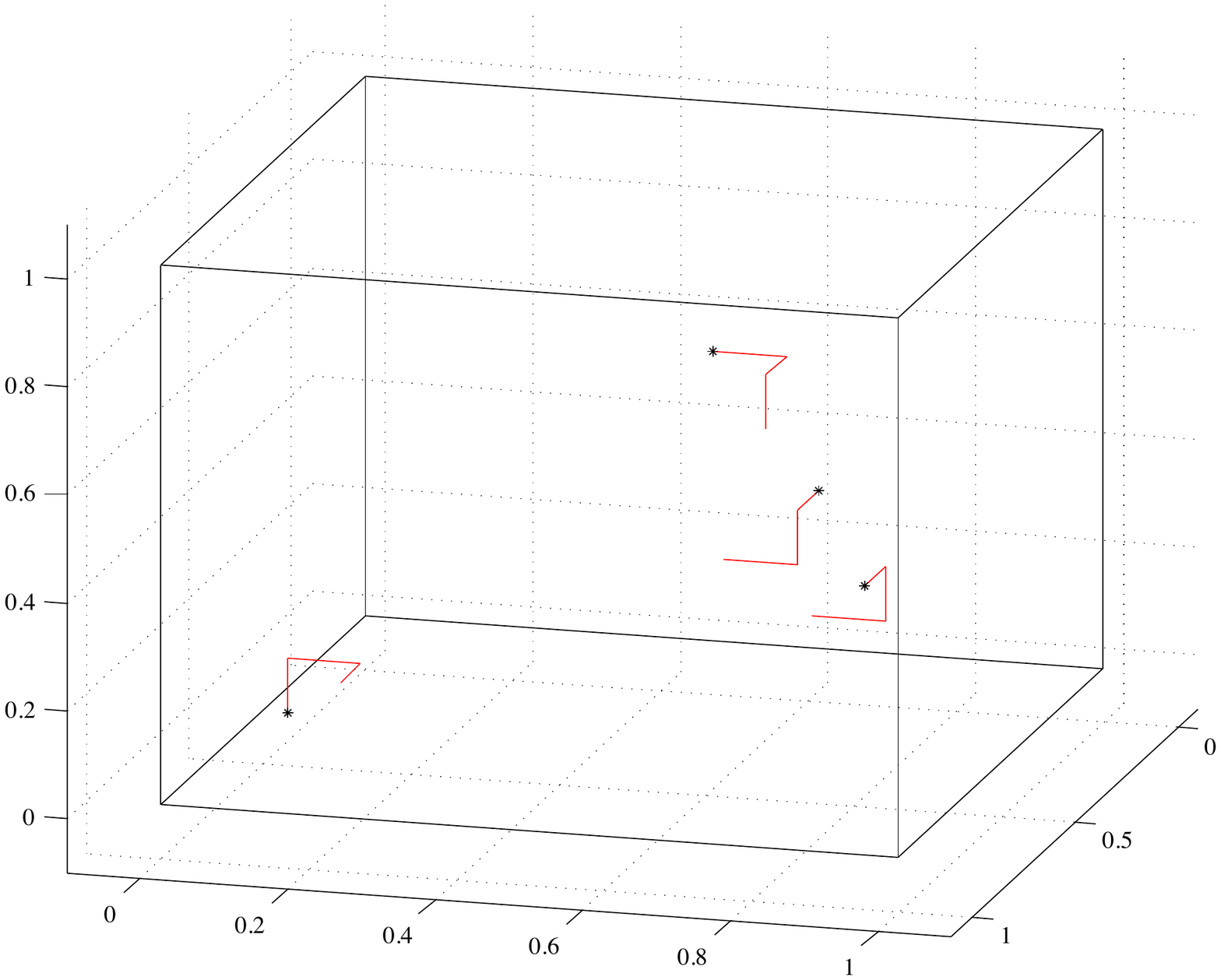}\hspace{\fill}
}\fi
\end{center}
\vspace*{-2cm}
\caption{Illustration of Morris elementary effects method ($d=3, r=4$). \label{fig:morris1}}
\end{figure}

If  points $\{\xi^i_n\}_{n=1, i=1}^{r,d}$ are chosen completely at random, the sensitivity analysis of a function of $d$ variables requires a total of $2dr$ evaluations of $f(\cdot)$.
The basic Morris scheme is a One-At-a-Time (OAT)   method that increases  efficiency with respect to  random sampling by using most evaluations of $f(\cdot )$ twice.
 It  relies on empirical moment estimates using $r$ samples of $\{d_i(\cdot)\}_{i=1}^d$ computed along $r$ randomly oriented paths $T_{d+1}$ along which each one of the $d$ coordinates is changed at a time, see Figure \ref{fig:morris1}.
 The total number of   evaluations  of $f(\cdot)$ is $r(d+1)$, which for large values of $r$ and $d$ may still be prohibitive.
Morris  clustered designs, see Section 5 in  \cite{morris:91}, improve on the efficiency of  these OAT designs by using each value of $f$  in the computation of more than two elementary differences. The simple paths $T_{d+1}$ are replaced by denser graphs that enable determination of $m\geq 1$ elementary differences along each direction.

For large values of $d$, Morris relies on a factorisation of the input space.
Let $Q_d$ be the $d$-dimensional unit hypercube, and factor $d=tq$ such that  $Q_d= Q_q^t$.  Let  $Y=\{\xi_1,\ldots , \xi_{|Y|}\}$ be a design  in $Q_q(\xi)$  that enables the determination of $m$ elementary effects along each direction. The full design  
$ \Xi = \bigcup_{j=1}^t Y_{(j)}$, where $Y_{(j)}$ is a replication of $Y$ along coordinates $X_{(j-1)q+1},\ldots, X_{jq}$, computes at least $m$ elementary effects along each  coordinate.

Although this idea is interesting, Morris' presentation is affected by a number of drawbacks.
In  \cite{morris:91} the smaller designs $Y\subset Q_q$ gather all $s\in Q_q$ with $\ell$ bits equal to one for all $\ell\in{\cal I}\subset\{0,\ldots, q\}$. Design multiplicity $m$ indirectly follows from choice of
 ${\cal I}$, but no guidelines on how this list should be chosen are provided, and actually,  since not all integers can be decomposed as the sum of a set of powers of two, not all multiplicities $m\leq 2^{d-1}$ can be obtained. Note also that  $d$  must not not prime and
\begin{equation}
d =tq \geq 2q_{\min}(m) = 2 \lceil \log_2(m)\rceil + 2, \qquad q\geq q_{\min}(m) = \lceil\log_2(m) \rceil+ 1 \label{eq:defqmin}
\enspace .
\end{equation}

 Verification of the properties of Morris' clustered designs is cumbersome and their optimality, as  it is recognised by the author, is not guaranteed. In fact, since  Morris designs are {not} necessarily connected -- they will be if  $q\in \cal I$  -- they are not natural candidates for optimality.

\subsection{Contributions}
The main result of the paper is  the explicit  presentation of a family of subgraphs of $Q_d$ that enable the computation of  a pre-specified number  $m$ of elementary effects for all $1\leq m\leq 2^{d-1}$.

\begin{definition}
A subgraph $S$ of $Q_d$  is a $(d,m)$-edge equitable design if and only if the number of edges of $S$ along each direction is exactly $m$.
\end{definition}

Figure \ref{fig:equitable} illustrates this definition. For each graph, edge colour indicates the direction of $Q_d$ along which the edge is aligned (we will use this colour code for arbitrary values of $d$). The number of edges of each colour is thus exactly equal to $m$ for  $(d,m)$-edge equitable graphs. The graph on the left is $(3,2)$-edge equitable, while the other two graphs are not edge equitable.

Although a vast literature characterising interesting families of subgraphs of the hypercube,  such as median and meshed subgraphs,  as well as on graph colouring problems, exists  in discrete mathematics, see e.g. \cite{Harary:88}, we could find no reference addressing this class of subgraphs, and their determination seems to be largely an open problem.

We present recursive procedures (Algorithms 1, 2 and 3) that generate $(d,m)$-edge equitable designs, overcoming most of the drawbacks of Morris' construction: {\em (i)} they are guided by the values of $m$ and  $d$, {\em (ii)} handle  generic values of $(d, m)$,  and {\em (iii)}   provably lead to equitable designs. 

We claim an additional contribution, that consists in  the exploitation of a convenient  polynomial representation of subgraphs of $Q_d$.  A related map between polynomials and  subgraphs of  $Q_d$, the log map,  has been used in \cite{Pistone:96}  to study the class of polynomial models identifiable by a design,  using computational commutative algebra. We believe that  the polynomial representation of subgraphs of $Q_d$ and, more importantly,  the exploitation of a suitably defined scalar product over polynomials for formal verification of several graph properties, without having to resort to intricate combinatorial arguments, is novel. In  particular, we are able to provide algebraic demonstrations for equitability (Theorem \ref{th:firstconst}), and derive explicit formulas for the size of our designs (Theorems \ref{th:size}, \ref{th:size2} and \ref{th:size3}).\\

As the paper shows, improved efficiency in the computation of the elementary effects by using clustered designs does not translate into better performance on the classification of input factors in Morris original method.  Definition of  tests adapted to the structured sampling implemented by clustered designs will be addressed in a forthcoming paper.

\section{Polynomial representation of subgraphs of $Q_d$}\label{sec:poly}

We  concentrate on subgraphs of the unit hypercube $Q_d=\{0,1\}^d$, i.e., the graph whose vertices are the points having coordinates $0$ or $1$ in $\mathbb{R}^d$, two points being  joined by an edge if only if they differ in exactly one coordinate. 
Given an ordering of the directions of $Q_d$, there is a bijection between its vertices  and the binary words of length $d$:
\[
\begin{array} {rcl} Q_d &\rightarrow &\{0,1\}^d\\
s &\hookrightarrow &\{s_i\}_{i=1}^d, s_i \in\{0,1\} \ .
\end{array}
\]
We define a $d$-edge-coloring of $Q_d$ by  stating that an edge joining two  points $s$ and $s'$ has color $i$ when $s_i\neq s'_i$ and $s_j= s'_j, j\neq i$. 

We associate to each $s\in Q_d$ a monomial $\mathcal{P}_s$
in the ring $\mathbb{R}[X_1,\ldots ,X_d]$ of the polynomials over the variables $X_1,\ldots ,X_d$:
\begin{equation*}
s=\{s_1,\ldots,s_d\} \longrightarrow \mathcal{P}_s(X_1,\ldots, X_d)=X_1^{s_1}\ldots X_d^{s_d}\enspace .
\end{equation*}
  The subgraph induced by a set $S\subset Q_d$ will be represented by the polynomial $\mathcal{P}_S=\sum_{s \in S}\mathcal{P}_s$. The empty set is represented by the zero polynomial. 
The set of the polynomials representing simple subgraphs of $Q_d$  will be denoted by $K_d$, and corresponds to the polynomials of degree at most $1$ in each variable having coefficients in $\{0,1\}$.

\subsection{Scalar product in $K_d$}

The set $K_d$ can be embedded in the algebra $\mathbb{R}[X_1,\ldots ,X_d]/\{X_i^2\equiv1,i=1\ldots d\}$   of the polynomials according to the equivalence relation induced by the equalities ${X_i^2\equiv1,i=1\ldots d}$.
This algebra is a vector space for which the set of monomials can be taken as a natural basis. By defining a scalar product  such that this basis is orthogonal, we endow $K_d$ with a structure that has several interesting properties   in term of the underlying subgraphs of $Q_d$.

\begin{definition}{}
We define the scalar product between monomials $\mathcal{P}_s,\mathcal{P}_{s^\prime}\in K_d$ as
\[ <\mathcal{P}_s,\mathcal{P}_{s^\prime}>=1_{s=s'}\enspace,\] and extend it naturally to  the entire $K_d$ by bilinearity
\[<\mathcal{P}_S,\mathcal{P}_{S^\prime}> = \sum_{s \in S,s \in S'}<\mathcal{P}_s,\mathcal{P}_{s^\prime}>, \qquad\mathcal{P}_S, \mathcal{P}_{S^\prime} \in K_d\enspace . \]

\end{definition}

\begin{figure}[h]
\begin{center}
\hspace{\fill}\includegraphics[width=.33\textwidth]{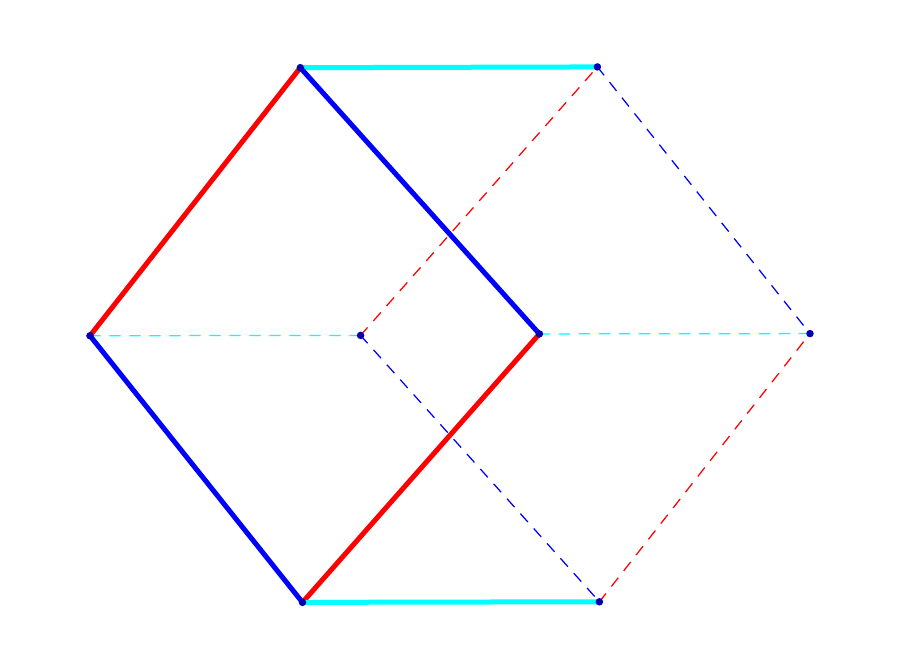}\hspace{\fill}\includegraphics[width=.33\textwidth]{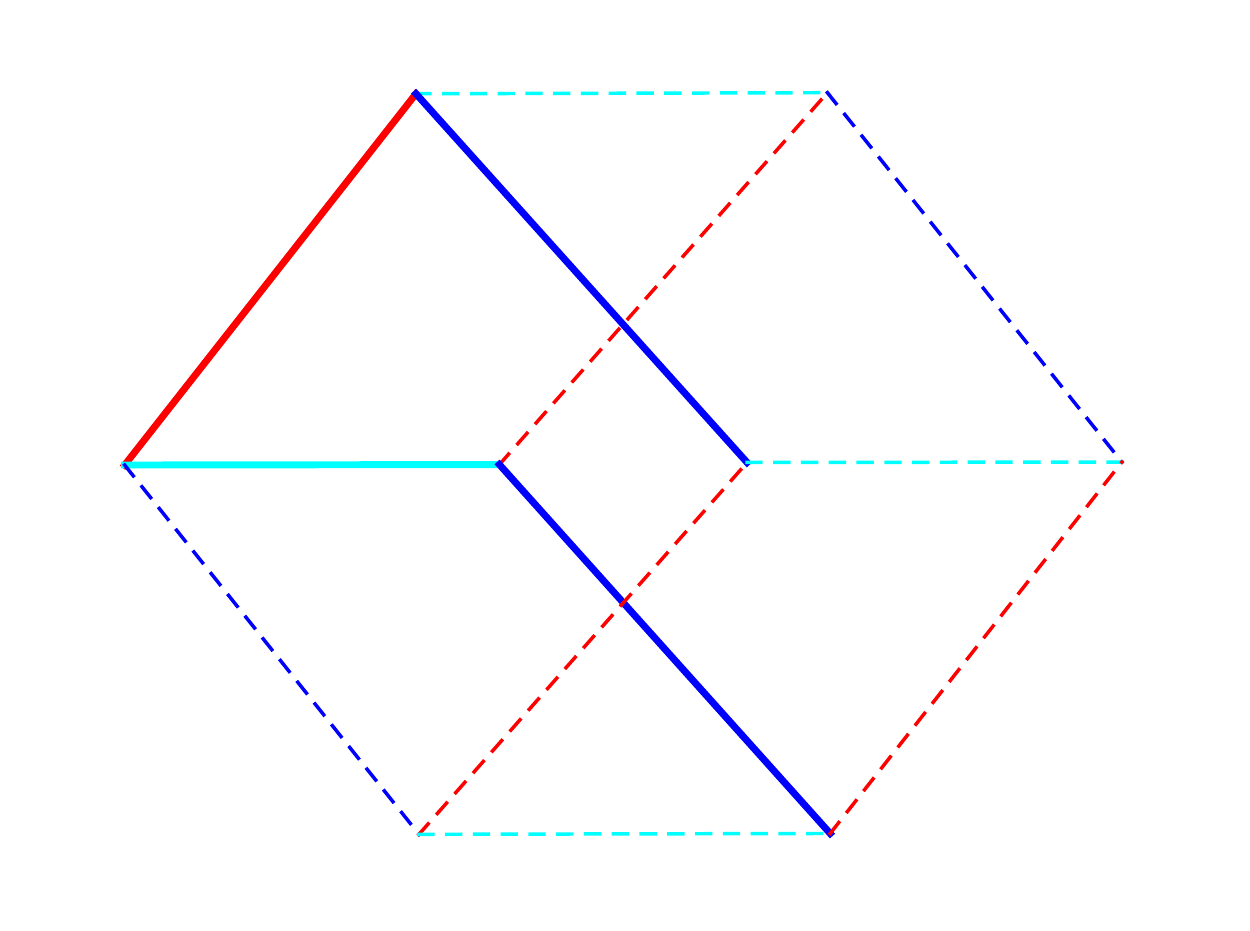}\hspace{\fill}\includegraphics[width=.33\textwidth]{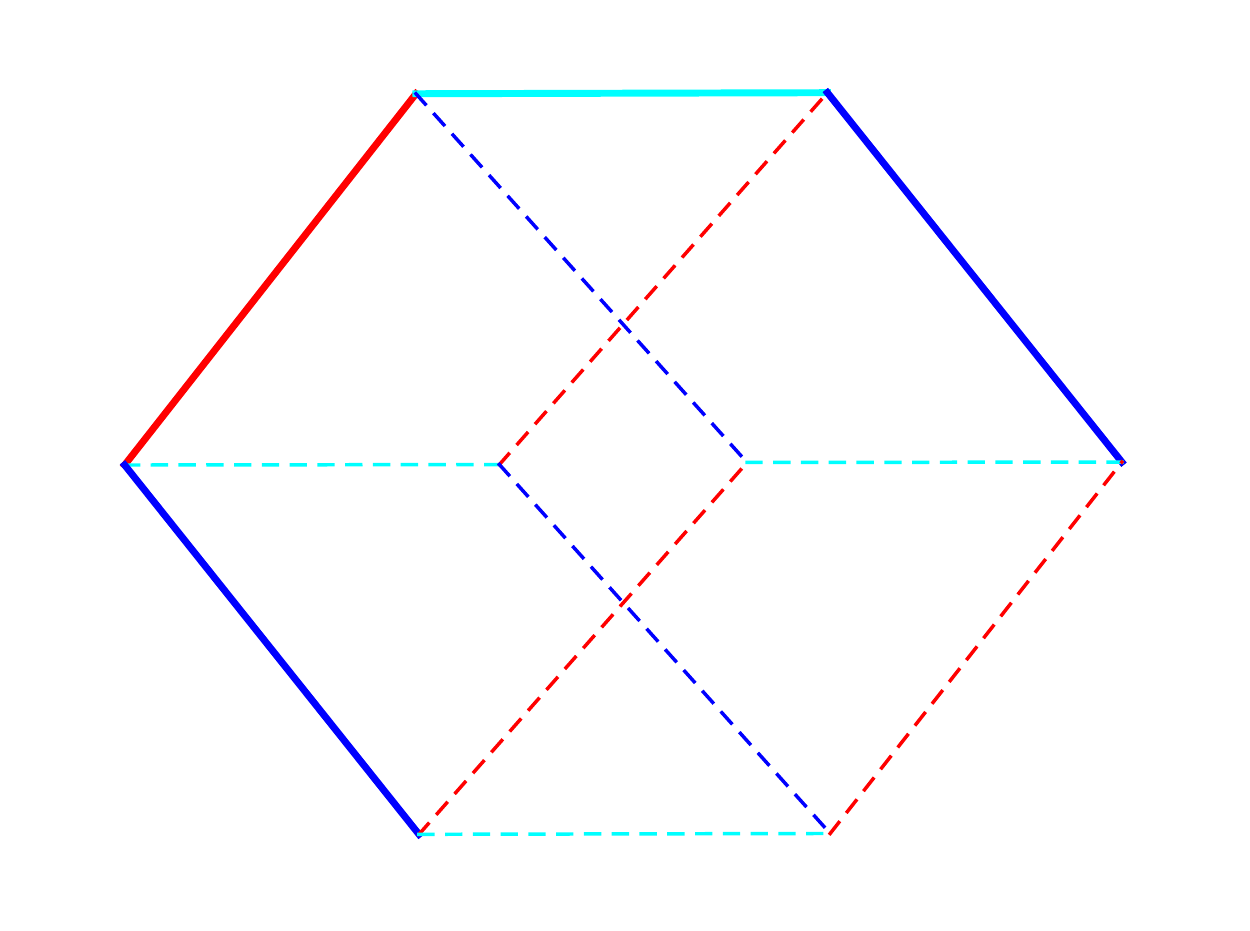}\\
\end{center}
\caption{Edge equitable (left) and non edge equitable (centre) graphs.  The colour of each edge indicates the direction  along which it is aligned. \label{fig:equitable}The graph on the right is the reflection of the graph on the left along $X_1$. \label{fig:reflect}}
\end{figure}
 
 \begin{lemma} The scalar product of two subgraphs of $Q_d$,  $S_1$ and $S_2$, is equal to the size of their intersection: $ \left< {\cal P}_{S_1} , {\cal P}_{S_2}\right>=\left| S_1  \cap S_2 \right|$. In particular,  $<{\cal P}_S,{\cal P}_S>=||{\cal P}_S||^2=|S|$.\label{fact:size}
\end{lemma}

\begin{lemma}\label{fact:mirror}
Let $s \in Q_d$ and $S \subset Q_d$. 
The subgraph  $S^\prime$  defined by
$
{\cal P}_{S^\prime} = {\cal P}_s {\cal P}_S
$
 is the reflection of $S$   along the directions present in $s$.\\ In particular, $X_i {\cal P}_S$ corresponds to the mirror of $S$ along direction $i$. 
\end{lemma}

Figure \ref{fig:reflect} illustrates Lemma \ref{fact:mirror}, showing ${\cal P}_S = 1+X_1+X_2+X_1X_3+X_2X_3$ and $X_1{\cal P}_S$. Multiplication by $X_1$ resulted in a reflection of $S$ along the red ($X_1$) direction.

\begin{lemma} For all  $s\in Q_d$,  $S, S^{\prime} \subset Q_d$
$<  {\cal P}_s  {\cal P}_S,  {\cal P}_s  {\cal P}_{S^\prime}> = < {\cal P}_S,  {\cal P}_{S^\prime}>\enspace .
$\label{lem:multip}
\end{lemma}
Using Lemmas \ref{fact:mirror} and \ref{fact:size} the following is immediate.

\begin{lemma} \label{fact:edges} The number $m_i$ of edges of $S\subset Q_d$ having color $i$ satisfies 
\begin{equation} <{\cal P}_S,X_i {\cal P}_S>=2m_i,\qquad i \in \{1,\ldots,d\} \enspace .
\end{equation}  \end{lemma}

\subsection{Problem (re)formulation}
Denote   by ${\cal E}^d_m$ the set of $(d,m)$-edge equitable polynomials. Using Lemma \ref{fact:edges},  
\begin{lemma}\label{def:equitable2}
${\cal P}_S \in  {\cal E}^d_m$  if and only if
\begin{equation}
<{\cal P}_S,X_i {\cal P}_S> = 2m,\qquad i \in \{1,\ldots,d\}\enspace .\label{eq:balanced}
\end{equation}
\end{lemma}
\begin{lemma} \label{lem:random} ${\cal E}^d_m$ is closed under multiplication by monomials:
\[
\forall s\in Q_d, \qquad {\cal P}_S\in {\cal E}^d_m \Rightarrow  {\cal P}_{S^\prime} = {\cal P}_s{\cal P}_S  \in {\cal E}^d_m\enspace .
\]
and under permutations of the coordinates of $Q_d$.
\end{lemma}

\begin{theorem}\label{th:complem} Let $S\subset Q_d$, and $\overline{S}$ denote the complement of $S$ in $Q_d$. 
\[{\cal P}_S \in {\cal E}^d_m \Rightarrow  {\cal P}_{\overline{S} }\in {\cal E}^{d}_{2^{d-1}+m-|S|}\enspace ,
\]
i.e., the complement of a $(d,m)$-edge equitable subgraph is an  $(d,m^\prime)$-edge equitable graph, with $m^\prime = 2^{d-1}+m-|S|$.
\end{theorem}
{\em Proof.}\\
Let ${\cal P}_S \in {\cal E}^d_m$ and compute
 $\left< {\cal P}_{\overline{S}}, X_i {\cal P}_{\overline{S}}\right>$
\begin{eqnarray*}
\left< {\cal S}_{\overline{S}}, X_i {\cal P}_{\overline{S}}\right> &=& \left< {\cal P}_{Q_d} - {\cal P}_{S}, X_i ({\cal P}_{Q_d} - {\cal P}_{S})\right>\\
&=& \left< {\cal P}_{Q_d} , X_i {\cal P}_{Q_d} \right> + \left< {\cal P}_{S}, X_i {\cal P}_{S}\right> - 2 \left< {\cal P}_{Q_d} , X_i {\cal P}_{S}\right>=2^d +2m -2 |S| = 2 m^\prime\ ,
\end{eqnarray*}
which is independent of $i$, completing the proof.

Theorem \ref{th:complem} is a first illustration of the power of the polynomial representation for establishing the properties of subgraphs of the hypercube. 

\section{Generation of  $(d,m)$-equitable subgraphs of  $Q_d$}
\label{sec:edge}

\subsection{Recursive graph composition} \label{sec:const1}

For every natural number $d$ and every integer $1 \leq m \leq 2^{d-1} $ the algorithm below produces a polynomial $G^d_m\in {\cal E}_m^d$.   

 {\bf Algorithm 1.} \mbox{\em Recursive definition of $G^d_m$.}\\
{\em Initialization} ($m=1$):
\begin{equation}
G^d_1 = 1+\sum_{i=1}^d X_i\enspace. \label{eq:init1}
\end{equation}
{\em Recursion}
\begin{align} \bullet \mbox{ For $m$   even, }&\qquad  G_m^d=(1+ X_1X_d) G^{d-1}_{^m/_2}\enspace .\label{eq:AlgoEvenM}\\
\bullet \mbox{ For $m$   odd,\ \ \,}&\qquad G^d_m=G^{d-1}_{^{(m-1)}/_2}+X_1X_d G^{d-1}_{^{(m+1)}/_2}\enspace .\label{eq:AlgoOddM}\end{align}

%

\begin{figure}
\begin{center}
\includegraphics[width=.7\textwidth]{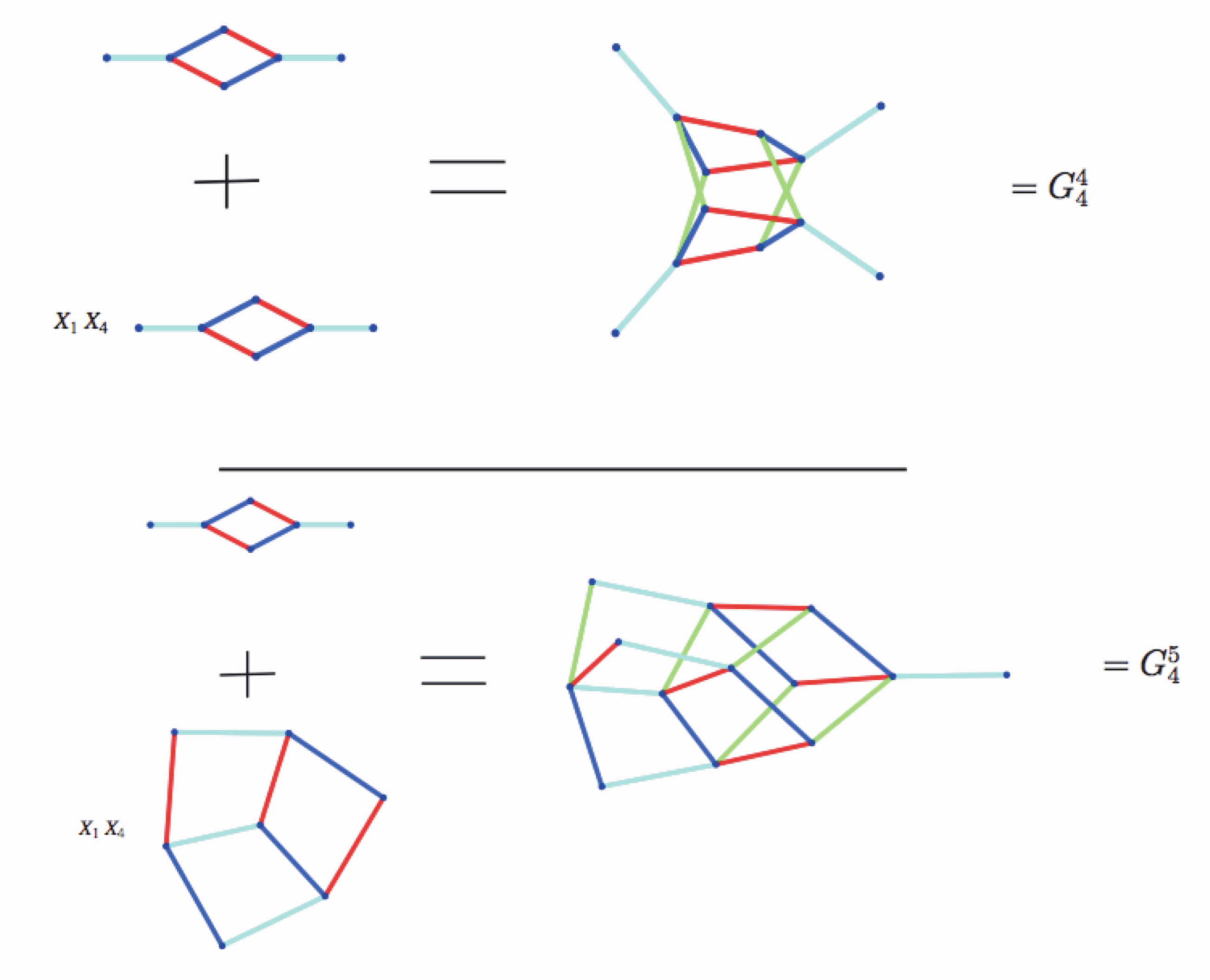} 
\end{center}
\caption{Construction of $G^4_4$ (top) and of $G_5^4$ (bottom). \label{fig:recur}}
\end{figure}

Figure \ref{fig:recur} 
illustrates the graph compositions of eq. (\ref{eq:AlgoEvenM})  ($m$ even) and eq. (\ref{eq:AlgoOddM})  ($m$ odd), respectively. Note that in the graphs displayed on the right the $m$ edges linking the two graphs on the left  are along the new dimension $X_4$   (green colour).
The solutions are  the composition of graphs  with  smaller values of $d$ and $m$, along a binary tree whose   leaves all have $m=1$. 

%

\begin{theorem}\label{th:firstconst}
For all $d\geq 1$, and all $1\leq m\leq 2^{d-1}$, the graphs $G^d_m$ defined by (\ref{eq:init1}) -- (\ref{eq:AlgoOddM}) are $(d,m)$-edge equitable. 
\end{theorem}

The proof of Theorem \ref{th:firstconst} is given in Appendix A, and is based on the Lemmas of Section \ref{sec:poly}.
Theorem \ref{th:firstconst} overcomes major limitations of Morris presentation, by defining a solution {\em (i)} for all pairs $(d,m)$ which is {\em (ii)} provably equitable.



\begin{theorem}\label{th:size}
For the graphs $G^d_m$ defined by (\ref{eq:init1}) -- (\ref{eq:AlgoOddM})
\begin{equation}
\left| G^d_m \right| = m(d-\kappa) +2^{\kappa+1} -m,\qquad \mbox{where }\kappa = \lfloor \log_2(m)\rfloor\enspace . \label{eq:sizeG}
\end{equation}
\end{theorem}

Demonstration of Theorem \ref{th:size} is trivial by verifying that (\ref{eq:sizeG}) is satisfied if we initialise with $\left| G^d_1 \right| =d+1$ the recurrence equations implied by (\ref{eq:AlgoEvenM}) and (\ref{eq:AlgoOddM})
\begin{align*}
\left| G^d_m \right| &= 2 \left| G^{d-1}_{^m/_2} \right|,\qquad \mbox{for $d$ even}\\
\left| G^d_m \right| &= \left| G^{d-1}_{^{m-1}/_2}\right| + \left| G^{d-1}_{^{m+1}/_2} \right|,\qquad \mbox{for $d$ odd}\ .\\
\end{align*}
%

\subsection{Improving efficiency by an alternative initialisation}\label{subsec:init2}\label{sec:const2}
Since  $\left|G^d_m\right|$ is recursively defined, decreasing size  for small values of $m$ will propagate to larger values of $m$.
We present now an alternative family of $(d,m)$-edge equitable graphs $H^d_m$,  $m\geq 2$, obtained by starting the recursion (\ref{eq:AlgoEvenM}) -- (\ref{eq:AlgoOddM}) at $m=4$. 
{The minimality of these graphs can be proved for $m=2$, and  has been checked numerically for $m=3, d\leq 5$.}\\

\noindent{\bf Algorithm 2.} \mbox{\em Recursive definition of $H^d_m$.}\\
{\em Initialization} ($m=2,3$)
\begin{itemize}

\item  For $m=2$ we distinguish the cases of even and odd $d$:  
\begin{align}
\mbox{when $d\geq 2$ is even } & \qquad H_2^{d} =H_2^{d-2}+ (X_{d-1}+X_d+X_{d-1}X_d),\\
\mbox{ when $d\geq 3$ is odd, } & \qquad H_2^{d}=H_2^{d-1}+ X_{1}X_{d} +X_{d-1}X_{d} \enspace .
\end{align}
\item For $m=3$, $\qquad H_3^{d}=1+X_1X_d+\sum_{k=1}^d X_k+\sum_{j=1}^{d-1}X_jX_{j+1}$.

\end{itemize}
{\em Recursion}: 
Apply eqs. (\ref{eq:AlgoEvenM}) -- (\ref{eq:AlgoOddM}) .\\
The size of these graphs satisfies the recursive equations
\begin{align}
 |H_2^d| & =\left\{ \begin{array}{ll}  |H_2^{d-2}| + 3(d-2),\qquad \hbox{if } d\hbox{ is even}\\
|H_2^{d-3}| + 5, \qquad \qquad \ \ \  \hbox{if } d\hbox{ is odd} \end{array}\right. \label{eq:rec2}\\
 |H_3^d| & = 1+ 2d \nonumber
\end{align}
By writing $m\geq 3$ as $m= 2p_2 + 3p_3, p_2,p_3 \in\mathbb{N}^0$,
where $p_2$ and $p_3$ are the number of leaves labeled $2$ and $3$, respectively, in the recursive decomposition of $m$ used in our algorithm, the following Lemma can be demonstrated:

\begin{lemma} 
\label{fact_p23}Let $k = \lfloor \log_2(m) \rfloor$, and write $m = 2^\kappa+2^{\kappa-1}+i, \in [-2^{\kappa-1}, 2^{\kappa-1}[$. Then the number of subgraphs $H^\star_2$ and $H^\star_3$ in the recursive composition (\ref{eq:AlgoEvenM})-(\ref{eq:AlgoOddM}) are, respectively,
\begin{equation*}
p_2 = \left\{ \begin{array}{ll} 2i, &\mbox{\rm if } i\geq0\\
-i, & \mbox{\rm  if } i < 0 \end{array} \right. , \qquad p_3 = 2^{\kappa-1} - \left| i \right|\ .
\end{equation*}
If $i<0$ all the subgraphs are in dimension $d-\kappa +1$, otherwise the subgraphs with $m=2$ are in dimension $d-\kappa $.
\end{lemma}

\begin{theorem}\label{th:size2}
Let $\kappa $ and $i$ be defined as in Lemma \ref{fact_p23}. 
The size of  $H^d_m$ is
\begin{eqnarray}\label{eq:sizeEqs}
|H^d_m| &=& c(m) +  \alpha(m) d, \\
 &&\alpha(m) = \left\{ \begin{array}{ll} i+2^{\kappa}, & i\geq 0\\
i/2+ 2^ \kappa, & i \leq 0 \end{array}\right. =
\left\{ \begin{array}{ll} m-2^{\kappa-1}, & i\geq 0\\
\frac{1}{2}\left( m+2^{\kappa-1}\right), & i< 0 \end{array}\right.\enspace ,\nonumber
\end{eqnarray}
where the term independent of $d$ is
\[
c(m) = \left\{ \begin{array}{ll}-m \left(\frac{1}{2} \left((-1)^{d-\kappa}+1\right)+ \kappa\right)+2^{\kappa-2} \left(3 (-1)^{d-\kappa}+2 k+9\right) , & i \geq 0\\ 
-\frac{1}{2} m \left(\frac{1}{2} \left((-1)^{d-\kappa}-1\right)+ \kappa\right)-2^{k-3} \left(-3 (-1)^{d-\kappa}+2 \kappa-9\right), & i< 0 \end{array}\right.
\]
\end{theorem}
Proof is simple by verifying the recursive equations (\ref{eq:rec2}).\\
For large values of $\kappa $, it can be shown that
\[
|H^d_m| \simeq \left\{ \begin{array}{ll}
(d-\kappa)(m-2^{\kappa-1}), & i \geq 0\\ 
\frac{1}{2}(d-\kappa)(m+2^{\kappa-1}), & i< 0 \end{array}\right.
\]
When $m=2^ \kappa +2^{\kappa-1}$ a simpler expression can be found:
\begin{align}
\left|H^d_{2^ \kappa +2^{\kappa-1}}\right| & = 2^ \kappa\left( (d-\kappa)+ \frac{3}{2}\right)\enspace .\label{eq:appm}
\end{align}

Theorem \ref{th:size2} completes the characterisation of the family of solutions $\overline{H}^d_{m^\prime} = \overline{H^d_m}$, in terms of their size, which is $|\overline{H}^d_{m^\prime}|=2^d - |H^d_m|$, and  of the value of $m^\prime$.

\subsection{Further improving economy by factoring the designs}\label{sec:const3}
Consider the case $m=2^ \kappa +2^{\kappa-1}$ when the simpler expression in (\ref{eq:appm}) holds. We can check that
\[
|H^{2d}_m| = 2|H^d_m| + 2^ \kappa\left(\kappa-\frac{3}{2}\right) > 2|H^d_m|\enspace ,
\]
i.e., the size of our designs grows supra-linearly in $d$. It can be checked that this is true for generic values of $m$.
 We improve the family of designs presented in the previous subsection by combining the factorisation approach used by Morris clustered designs with the generic solution $H^d_m$ presented in the previous section.

Remember the definition of $q_{\min}(m) = \lceil \log_2(m)\rceil  +1$, see equation (\ref{eq:defqmin}), as the dimension of the smallest hypercube that can contain  $m$ edges along each direction, and for a given pair $(d,m)$ write $d$ as 
$
d = c\cdot  q_{\min}(m) + t^\prime 
$,
such that
\begin{equation}
d = (c-1) q_{\min}(m) + t, \qquad t \in \{q_{\min}(m),\ldots, 2q_{\min}(m)-1\}\enspace .\label{eq:d_decomp}
\end{equation}
It is easy to check that $q_{\min}(m)= \kappa +1$, where $\kappa $ is the parameter in Lemma \ref{fact_p23}.
For $d\geq2q_{\min}(m)$ designs more efficient than those presented in section \ref{subsec:init2} can be obtained by placing $c-1$ copies of $H^{q_{\min}(m)}_{m}$ in disjoint $q_{\min}(m)$-dimensional subspaces of $Q_d$, and adding a $H^t_m$ design covering the remaining directions.
In the following we will often omit indication of the dependency on $m$, using the simpler notation $q_{\min}$.

\noindent{\bf Algorithm 3.} \mbox{Definition of $M^d_m$.}\\
\begin{equation}\label{eq:GFact}
M_{m}^d = 1 + \sum_{j=1}^{c-1}\left[ \mbox{Shift}_{jq_{\min}}\left(H^{q_{\min}}_m\right)-1 \right]+ \mbox{Shift}_{(c-1)q_{\min}}H^t_m\enspace ,
\end{equation}
where Shift$_k({\cal P})$ operates over the coordinates of the polynomial ${\cal P}$:
\[
\mbox{Shift}_k\left({\cal P}(X_{i_1}, \ldots, X_{i_n}) \right)= \left({\cal P}(X_{i_1+k}, \ldots, X_{i_{n}+k}) \right)\enspace .
\]

For $d=q_{\min}$, 
\[
\left| H^{q_{\min}}_m\right| =\left| H^{\kappa+1}_m\right| = \left\{ \begin{array}{ll} 
m+2^{\kappa-1}, &i \geq 0\\
\frac{1}{2}(m+ 2^{\kappa-1}) + \frac{3}{2}2^{\kappa}, & i< 0
\end{array} \right.
\]

\begin{theorem}\label{th:size3}
The size of the designs defined by (\ref{eq:GFact}) is
\begin{align*}
\left| M_{m}^d\right| &= 1+ (c-1) |H^{q_{\min}}_m|+ | H^t_m|\\
&= \left\{ \begin{array}{ll}
\left(\lfloor \frac{d}{\kappa+1}\rfloor-1\right) \left(m+2^{\kappa-1}\right) +| H^t_m|, & i\geq 0\\
\left(\lfloor\frac{d}{\kappa+1}\rfloor-1\right) \left(\frac{1}{2}(m+ 2^{\kappa-1}) + \frac{3}{2}2^{\kappa}\right) + | H^t_m|, & i<  0
\end{array} \right.
\end{align*}
where $c$, $t$ and $q_{\min}=\kappa+1$ are defined in eq. (\ref{eq:d_decomp}), and $| H^t_m|$ is given by Theorem \ref{th:size2}.
\end{theorem}

\begin{figure}[h]
\begin{center}
\if0\bw
{
\includegraphics[width=.7\textwidth]{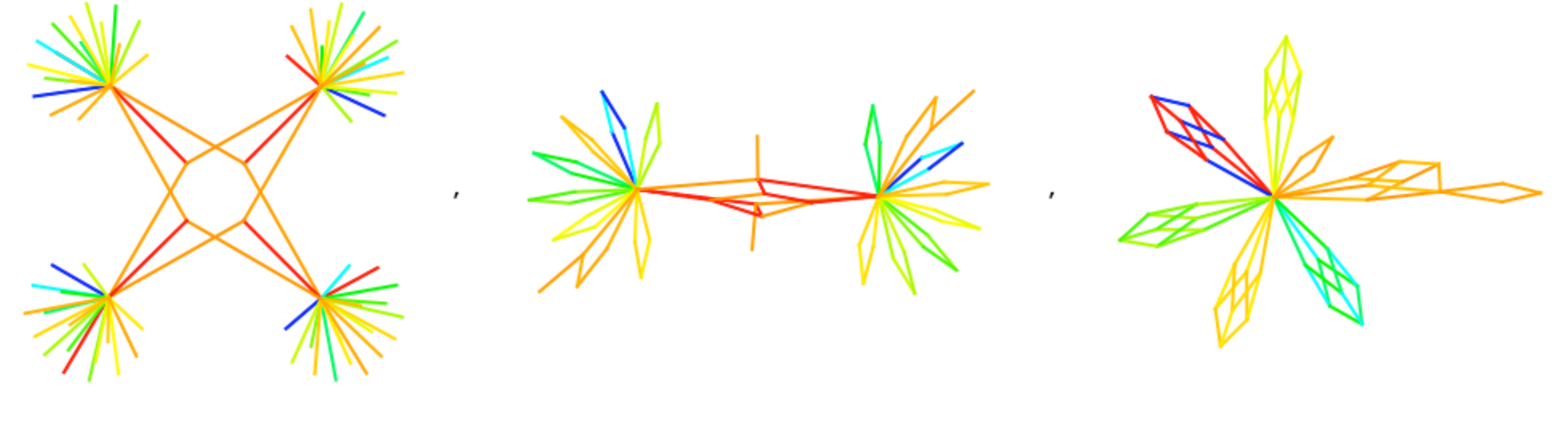}
}\fi
\if1\bw
{
\includegraphics[width=.8\textwidth]{Factored19bw}
}\fi
\end{center}
\caption{The three families of edge equitable graphs: $G^{19}_5$ (left), $H^{19}_5$ (centre) and $M^{19}_5$ (right). Sizes are 76, 60 and 49, respectively. \label{fig:allfamilies}}
\end{figure}

Figure \ref{fig:allfamilies} shows the three families of graphs for $d=19$ and $m=5$, an example of a situation for which Morris' construction is not defined. Note the remarkably different graphs topologies, as well as the decreasing size: $76=|G^{19}_5| > |H^{19}_5|=60 > |M^{19}_5|=49$.

We remark that for the values of $d$ and $m$ for which Morris designs are fully described in his paper, $M^d_m$ is a perfectly equitable design of the same size, but our our construction is defined for all pairs $(d,m)$.

\subsection{Economy}
Morris characterised efficiency of a design $S$ as the ratio of the total number of elementary effects that can be computed using $S$ to its size. We adhere to his definition.
\begin{definition}
Let $S\in {\cal E}^d_m$. The economy of $S$ is 
\begin{equation}
\Gamma(S) = \frac{md}{|S|}\enspace .\label{eq:economy}
\end{equation}
\end{definition}

\begin{lemma} 
$ |G^d_m| \geq |H^d_m|\geq |M^d_m|, \qquad \Gamma(G^d_m) \leq \Gamma(H^d_m)\leq \Gamma(M^d_m)$.
\end{lemma}

\begin{figure}
\begin{center}
\if0\bw
{
\hspace{0\textwidth}$\scriptsize \Gamma$\\
\includegraphics[width=.45\textwidth]{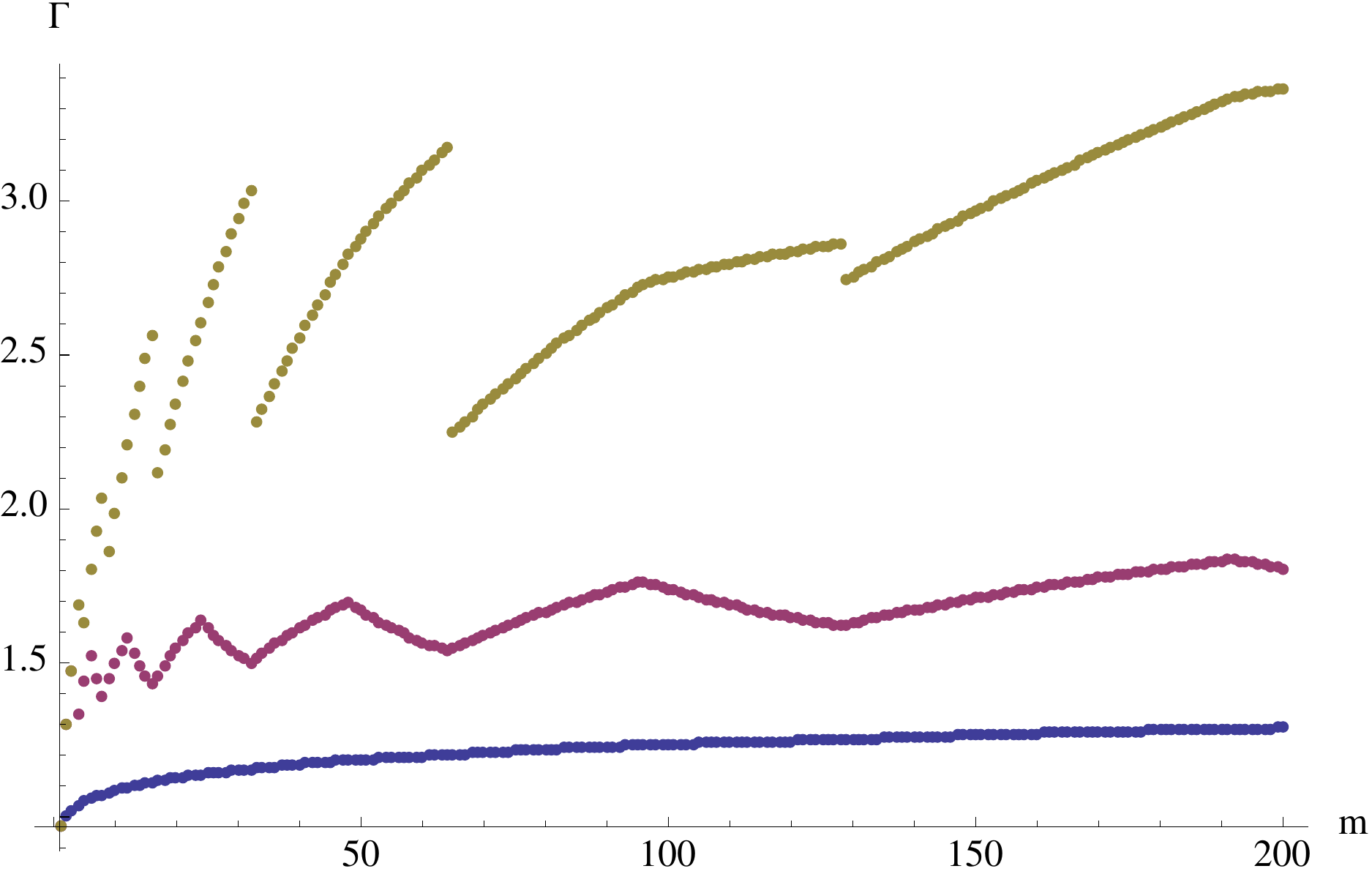}
\ \ \includegraphics[width=.45\textwidth]{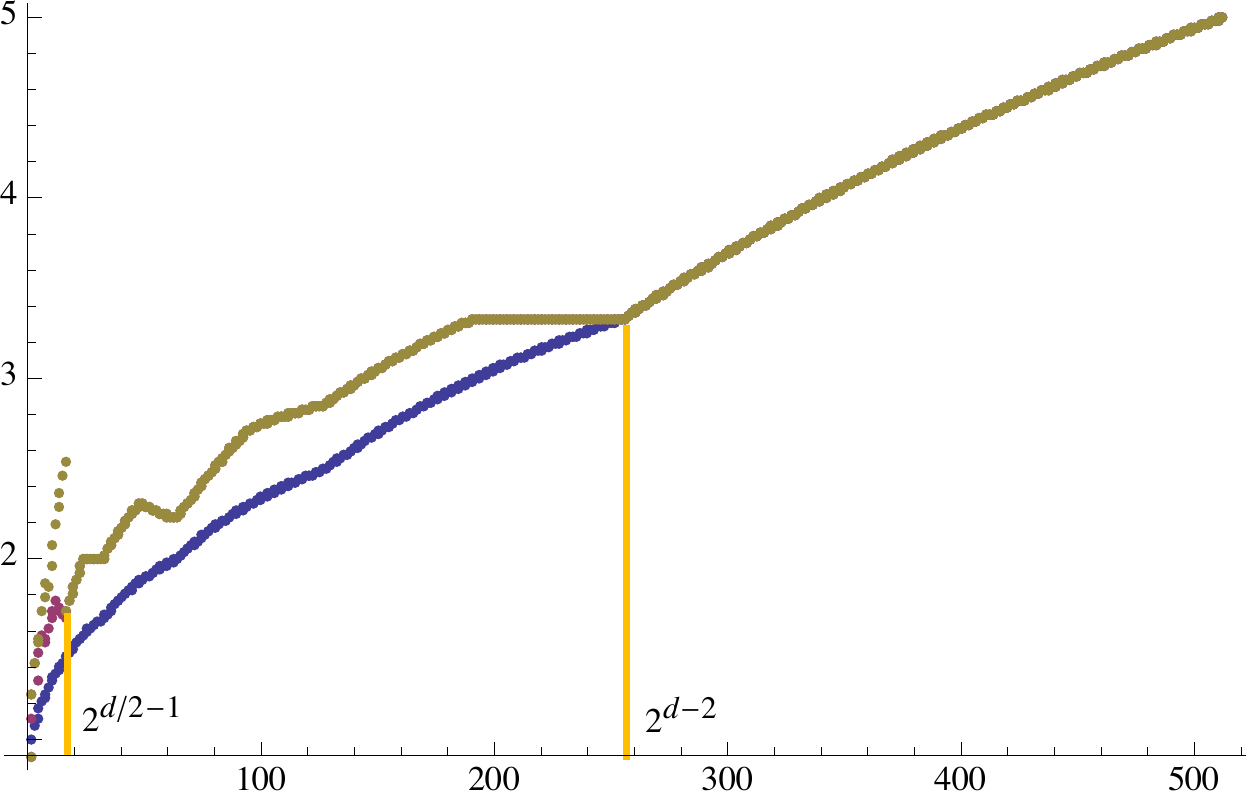}\ \ $m$
}\fi
\if1\bw
{
\hspace{0\textwidth}$\scriptsize \Gamma$\\
\includegraphics[width=.45\textwidth]{Economybw.pdf}
\ \ \includegraphics[width=.45\textwidth]{EconomyAllbw.pdf}\ \ $m$
}\fi

\end{center}
\caption{Economy ($\Gamma$) of designs $G^d_m$ (lower curves), $H^d_m$ (middle curves) and $M^d_m$,  (upper curves), $m\leq 200, d=30$ (left) and complete range of $m$, $d=10$ (right). \label{fig:economy}}
\end{figure}

 Figure \ref{fig:economy} confirms this Lemma:  the economy of $H^d_m$  is framed by the economies of $G^d_m$ (below) and of $M^d_m$ (above).  These plots confirm that  factorisation leads to a significant improvement, nearly doubling economy for small values of $m$.
 
 Note (see right plot, where entire range of $m$ for $d=10$ is plotted) that all three curves come together at an economy of $d/2$ for the upper limit  of $m=2^{d-1}$, i.e, when the hypercube becomes complete. The point at which all curves merge is $m=2^{d-2}$: for $m\geq 2^{d-2}$, $(d,m)$-edge equitable solutions are ``unique,'' in the sense that they all correspond to the deletion of $2^{d-1}-m$ non-adjacent points of $Q_d$. 
We can see that the middle curve ($H^d_m$) rapidly coincides with the (upper) curve for the factored design ($M^d_m$): since factored designs exist only if $d\geq 2q_{\min}$, i.e., for $m\leq 2^{(d/2-1)}$, from this point onwards the green and red curves are indistinguishable.

\begin{lemma} When $d\rightarrow \infty$
\begin{align*}
\Gamma^\infty(G^d_m) &= \lim_{d\rightarrow\infty} \frac{md}{|G^d_m|} =1\\
\Gamma^\infty(H^d_m) &= \lim_{d\rightarrow\infty} \frac{md}{|H^d_m|} = \frac{m}{\alpha(m)}= 
\left\{ \begin{array}{ll} \frac{1}{1-2^{\kappa-1}/m}, & i \geq 0\\ 
\frac{2}{1+2^{\kappa-1}/m}, & i < 0 \end{array}\right., \qquad\frac{4}{3} \leq \Gamma^\infty(H^d_m)  \leq \frac{3}{2}\\
\Gamma^\infty(M^d_m) &= \lim_{d\rightarrow\infty} \frac{md}{|M^d_m|}, \qquad \qquad \frac{m^2}{2m-1} \leq \Gamma^\infty(M^d_m) \leq  \frac{m^2}{m-1} 
\end{align*}
\end{lemma}
The expressions above follow from the definition of economy and the expressions for the size of the designs. Note that $\Gamma^\infty(G^d_m)$ is bounded below by $\Gamma^\infty(H^d_2)={^4}/_3$ and above by $\Gamma^\infty(H^d_3)={^3}/_2$. It is easy to check that $\Gamma^\infty(G^d_m)=\Gamma^\infty(G^d_1)=1$, showing that the economy of our recursively defined solutions is bounded by the economy of their initialisations.

\section{Sensitivity analysis}
\label{sec:applic}

For One-At-a-Time designs, the elementary effects $d_i$ can be computed incrementally as $f(\cdot)$ is evaluated at consecutive points of the design. This is no longer the case for $m>1$. 
We indicate below how the polynomial representation  can be exploited to identify the $m$ pairs of points $\{(i_1^{(\ell)},i_2^{(\ell)})\}_{\ell=1}^m$  involved in the computation of the $m$ elementary effects along direction $i$. Let ${\cal P}$ be the polynomial representation of the design and $n$ be its size (the number of terms in ${\cal P}$). Consider an ordering of the monomials of $\cal P$, such that
\[
{\cal P} = \sum_{p=1}^n {\cal P}_{s_p}, \qquad s_p \in Q_d\enspace ,
\]
and let $f_{\cal P}$ denote the vector of valuations of $f(\cdot)$: $[f_{\cal P}]_p= f(s_p), p=1,\ldots, n$. Define the $d$ upper-triangular matrices
\[
\left[E_i \right]_{(p,q)} = \left\{
	\begin{array}{ll} 
		(1-2 [{s}_p]_i)\left< {\cal P}_{s_p}, X_i {\cal P}_{s_q}\right>,& \enspace 1 \leq q < p \in \{1,\ldots, n\}, i\in\{1,\ldots, d\}\\
0,& \mbox{ otherwise}\end{array}\right.
\]
There is at most one non-zero entry in each line of $E$. 
Assuming that ${\cal P}$ is a $(d,m)$-edge equitable design, there are exactly $m$ non-zero elements $[E_i]_{(i_1^{(\ell)}, i_2^{(\ell)})}= \pm 1, \ell=1,\ldots, m$, that indicate the pairs of points of the design  that enable the computation of the $m$ elementary effects $d_i$, which are the $m$ non-zero entries of ($\mathbf{1}$ is the $d$-dimensional vector of 1's)
\[
 d_i  = \frac{1}{\Delta}J_i f_{\cal P},\qquad J_i =  - \mbox{diag}\left( E_i \mathbf{1} \right) + E_i \enspace .
\]
Sample averages can be computed (remember there are only $m$ non-zero values if $d_i$) as
\[
\mu_i =  \frac{1}{\Delta m} \mathbf{1}^T J_i f_{\cal P}, \qquad \mu^\star_i =  \frac{1}{\Delta m} \mathbf{1}^T \left| J_i f_{\cal P}\right|, \qquad 
i=1,\ldots, d\enspace .
\]

Consider the following example in ${\cal E}^4_2$:
\[ {\cal P} = 1 + X_1 + X_2 + X_1X_2 + X_1X_2X_3 + X_1X_2X_4 + X_1X_2X_3 X_4
\]
For this graph, $n=7$, and consider that the nodes are listed by order. Consider direction $X_2$, for which the non-zero elements of  $J_2$ are $[J_2]_{1,3} = [J_2]_{2,4} =1$, and $[J_2]_{1,1} = [J_2]_{2,2} = -1$, and thus
and 
\[ d_2^T = \frac{1}{\Delta}\left\{ \begin{array}{ll}
0, & i\not\in \{1,2\} \\
\left[f_{\cal P}\right]_3 - [f_{\cal P}]_1, & i=4\\
\left[f_{\cal P}\right]_4 - [f_{\cal P}]_2, & i= 6
\end{array} \right. \enspace .
\]

 Morris Elementary Effects method is based on a set of elementary effects computed along $r$ random perturbations of a basic design $\cal P$.
Using Lemma \ref{lem:random}, random versions of a design represented by polynomial $\cal P$ can be obtained as
\[
{\cal P}^{(j)} = s^{(j)} {\cal P}(X^{(j)} + \pi^{(j)} (X)), \qquad j=1,\ldots, r\ ,
\]
where $\{s^{(j)}\}_{j=1}^r$ are independent and  uniformly drawn in $Q_d$, $\{X^{(i)}\}_{j=1}^r$ are independent and uniform in $\cal A$ and $\{\pi^{(j)}\}_{j=1}^r$ are independent random permutations of $\{1,\ldots, d\}$.

\section{Numerical application}
\label{sec:numerical}
We illustrate in this section the application of the designs presented in the previous sections, considering the same function as used in the original publication \cite{morris:91}.
\begin{equation}
f(x) = \beta_0 +\sum_{i=1}^{20}\beta_i w_i + \sum_{i<j}^{20} \beta_{ij} w_i w_j + \sum_{i<j<l}^{5} \beta_{ijl} w_i  w_j w_l + \sum_{i<j<l<s}^{4} \beta_{ijls} w_i w_j  w_l w_s
\end{equation}
where $w_i=2X_i-1, i\in\{1,2,4,6,8,\ldots , 20\}$ and $w_i = 2.2 X_i/(X_i+0.1) -1, i\in\{3,5,7\}$. Coefficients $\beta_i$ are as follows:
\begin{align*}
\beta_i = 20, &\qquad i\in\{1,\ldots, 10\},\qquad\ \  \ \, \beta_{ij} =-15, &\qquad i,j\in\{1,\ldots,6 \}\\
\beta_{ijl} =-10, &\qquad i,j,l \in\{1,\ldots, 5\},\qquad \beta_{ijls} =5, &\qquad i,j,l,s \in\{1,\ldots, 4\}.
\end{align*}
All remaining 1$^{st}$ and 2$^{nd}$ order coefficients are independent realisations of a standard normal distribution, 
$
\beta_i\sim {\cal N}(0,1),\  i\not\in \{1,\ldots, 10\}, \beta_{ij} \sim {\cal N}(0,1),\  i,j\not\in i,j\in\{1,\ldots,6 \}$.
For this function the relevant classes of input factors are  
\begin{align*}
{\cal C}_0 & = \{ 11,\ldots, 20\},\qquad
{\cal C}_1  =\{8, 9, 10\},\qquad
{\cal C}_2  =\{1,\ldots, 7\}\ .
\end{align*}

We apply Morris test for $m = 4$ and $r=3$, leading to a total of number of elementary effects per direction equal to 12.  The total number of evaluations of $f(\cdot)$ with this degree of multiplicity is  $ n_{4} = 147$, while the computation of the same number of elementary effects with the standard designs ($m=1$) requires $n_1 = 12 (d+1) = 252$, i.e., almost two times more.

\begin{figure}
\hspace{4cm}$\sigma_i$
\begin{center}
\includegraphics[width=.5\textwidth]{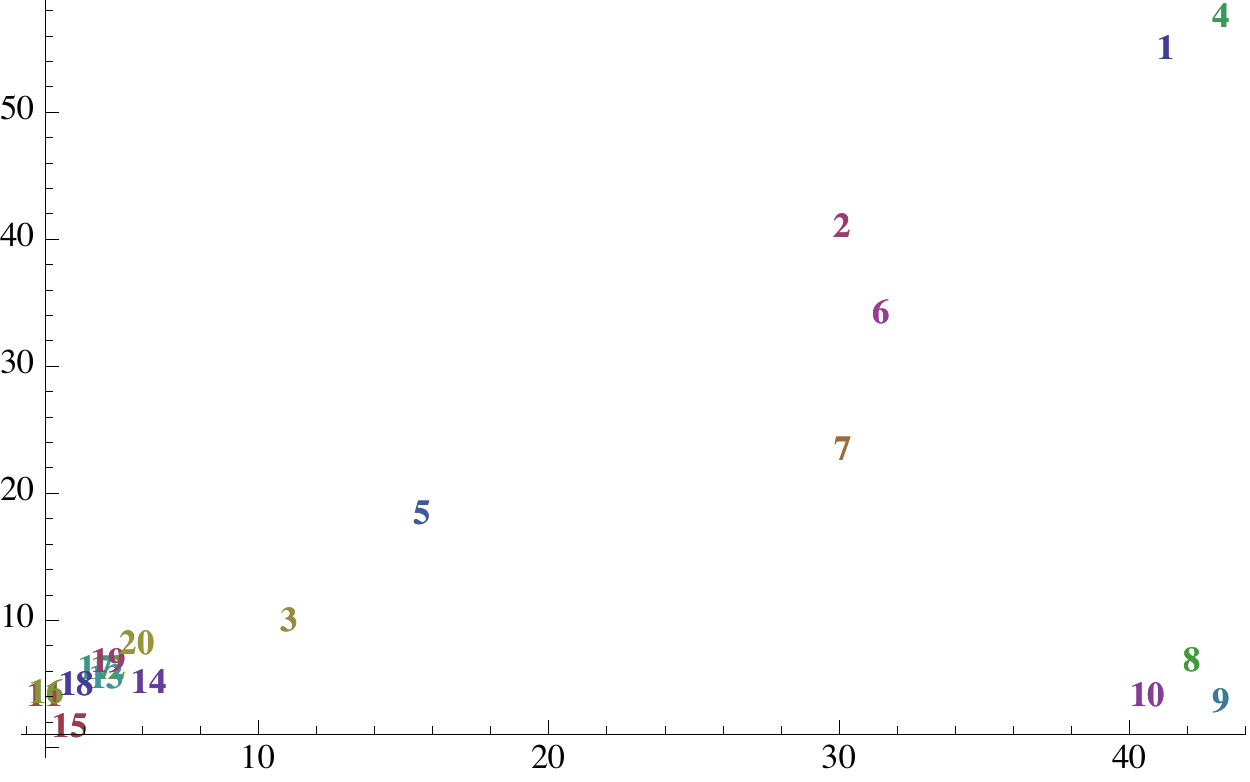}\ $\mu_i^\star$
\vspace{-4.4cm}
\setlength{\unitlength}{.9cm}
\begin{picture}(10,5)(-.5,0)
\qbezier(1,.2)(.2,.1)(1,1)
\qbezier(1,.2)(1.9,.3)(2,.7).
\qbezier(.9,1.)(2,1.9)(1.9,.7)
\put(1,1.5){${\cal C}_0$}
\qbezier(6.5,.8)(6.7,-.5)(7.7,.5)
\qbezier(6.5,.8)(6.4,2)(7,1.6)
\qbezier(7.7,.5)(8.1,1.1)(7.,1.6)
\put(8.5,1.8){${\cal C}_1$}
\qbezier(2,2.)(4.5,5)(8.,5.5)
\qbezier(2,2)(1,.7)(3,1)
\qbezier(3,1)(7,1.7)(8.3,4)
\qbezier(8.,5.5)(9.3,5.6)(8.3,4)
\put(5.5,5.2){${\cal C}_2$}
\end{picture}
\end{center}
\caption{Morris statistics ($m=4$) for $X_i, i=1,\ldots, 20$. Position of the label $i$  indicates the observed $(\mu_i^\star,\sigma_i)$. \label{fig:onerun}}
\end{figure}
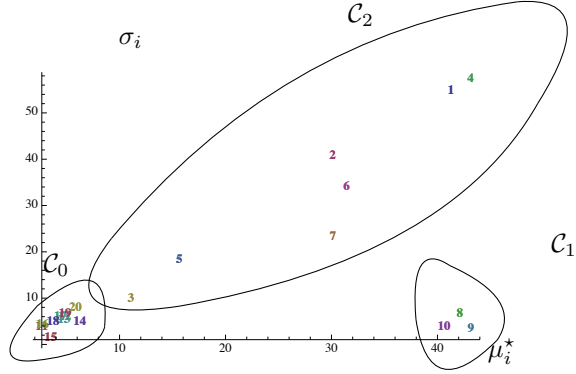

Figure \ref{fig:onerun} shows the statistics observed in one run of the test. The estimates of the variances $\sigma^2_i$ have been corrected to take into account the clustered nature of designs (see \cite{Skinner:81} for details). The three distinct classes are well identified, although some class ${\cal C}_2$ input factors, like $X_3$, come close to the  ${\cal C}_0$ region. This tendency to wrongly classify non-linear/mixed effects, which can occasionally be classified as linear or negligible, has ben recognised before, see \cite{campo:11}. In a subsequent paper we will fully address the study of Morris elementary method under clustered designs.

\section{Conclusions and further work}
\label{sec:conc}

The paper presents a complete and constructive definition of Morris clustered designs, that we designate by edge equitable designs.  The algorithms presented are based on a polynomial representation of subgraphs of the hypercube that enables simple algebraic manipulation of the graphs and determination of their properties. These algorithms overcome some limitations of the original presentation: we provide recursive algorithms that enable the construction of equitable graphs with arbitrary multiplicity $m$ for all dimensions $d$ of the input space.
 The results are novel, and we are not aware of a formal study of this class of graphs in the literature. 
 
 Some extensions are possible. Our designs are not minimal, and the determination of minimal edge-equitable graphs remains an open problem. Our designs are subsets of the hypercube. The approach based on a polynomial representation of graphs may be extended  to define equitable graphs over finite $d$-dimensional grids, recursively generated as  the iterated product of a basic finite set $S$: $
{\cal S}_d = {\cal S}_{d-1} \times S$, opening the way to computation of higher order derivatives. We are currently working in this direction.

Numerical studies show that there is a tradeoff between computational efficiency and discrimination power of the original Morris test when an increasing multiplicity $m$ is used. However, use of $m>1$ enables the definition of different kinds of tests, that will detect not just a large variability of the elementary effects across the entire domain of $f$, but how much their distribution changes over disjoint neighborhoods of the input space.  
This idea will be explored in future studies.

\appendix
\section*{Appendix A. Demonstration of Theorem \ref{th:firstconst}}
\label{sec:appendix}
We consider separately the cases of odd and even $m$.
\begin{itemize}
\item $m$ even\\
Assume that $G_{^m/_2}^{d-1}\in {\cal E}^{d-1}_{^m/_2}$, i.e., 
\[<G^{d-1}_{^m/_2},X_iG^{d-1}_{^m/_2}>= m, \qquad i\in\{1,\ldots, d-1\}\enspace .\] 
Since $G^{d-1}_{^m/_2} \in K_{d-1}$,
\[<G^{d-1}_{^m/_2},X_iX_dG^{d-1}_{^m/_2}>=0,\qquad \forall i\neq d\enspace .
\] 
 It follows immediately that, for $G^d_m$ defined by (\ref{eq:AlgoEvenM})
\[ <G_m^d,X_i G_m^d>=\left\{ \begin{array}{ll}
<G^{d-1}_{^m/_2},X_iG^{d-1}_{^m/_2}>+\\
\qquad <X_1 X_dG^{d-1}_{^m/_2},X_i X_1X_dG^{d-1}_{^m/_2}>, \mbox{ if } i< d\\
<G^{d-1}_{^m/_2}, X_1  G^{d-1}_{^m/_2}>+\\
\qquad <X_1 X_dG^{d-1}_{^m/_2},X_1 G^{d-1}_{^m/_2}>,\qquad \ \ \mbox {if } i =d \end{array} \right.
\enspace .\]
Each term in each branch of the right-handside of this equation is equal to $m$, demonstrating that $G_m^d\in{\cal E}^d_m$.

\item $m$ odd\\

 Assume  that $G^{d-1}_{^{m-1}/_2}\in{\cal E}^{d-1}_{^{m-1}/_2}$ and $G^{d-1}_{\frac{m+1}{2}}$. Then, for $G^d_m$ defined by (\ref{eq:AlgoOddM})
  \begin{align*}
  <G^d_m,X_i G^d_m> &= <G^{d-1}_{^{m-1}/_ { 2}},X_i G^{d-1}_{^ {m-1}/_{ 2}}> 
  +<X_1X_d G^{d-1}_{^ {m+1}/_{ 2}},X_i G^{d-1}_{^ {m-1}/_ { 2}}> \\
 & +<G^{d-1}_{^ {m-1}/_{ 2}},X_i X_1X_d G^{d-1}_{^ {m+1}/_{ 2}}> 
  + <X_1X_d G^{d-1}_{^ {m+1}/_ { 2}},X_i X_1X_d G^{d-1}_{^ {m+1}/_{ 2}}> \\
  &=\left\{ \begin{array}{ll}
<G^{d-1}_{^{m-1}/_{ 2}},X_i G^{d-1}_{^ {m-1}/_{ 2}}> +\\
\qquad + < G^{d-1}_{^ {m+1}/_{ 2}},X_i  G^{d-1}_{^ {m+1}/_{ 2}}> , \qquad\ \ \ \ \ \mbox{ if } i< d\\
2 <G^{d-1}_{^{m-1}/_{2}}, X_i X_1X_d G^{d-1}_{^{m+1}/_{2}}>,\qquad \ \ \mbox {if } i =d \end{array} \right.\\
&=\left\{ \begin{array}{ll}
(m-1)+(m+1) = 2m , \qquad\ \ \ \mbox{ if } i< d\\
2 <G^{d-1}_{^{m-1}/_{2}}, X_1 G^{d-1}_{^{m+1}/_{2}}>,\qquad \qquad \ \mbox {if } i =d \end{array} \right.
\end{align*}

Construction (\ref{eq:AlgoOddM}) will thus lead to a $(d,m)$-edge equitable subgraph iff our family of solutions verifies also the following additional condition
\begin{equation}
<G^{d-1}_m, X_1 G^{d-1}_{m+1}> =2m+1,\qquad \forall m\enspace .\label{eq:AlgoCondition}
\end{equation}

Assume that $(m-1)/2 = 2k$. Using equations (\ref{eq:AlgoEvenM}) and (\ref{eq:AlgoOddM}),
\begin{align*}
<G^{d}_{2k}, X_1 G^{d}_{2k+1}> &= <G^{d-1}_{k}, X_1 G^{d-1}_{k}>+<G^{d-1}_{k}, X_dG^{d-1}_{k+1}>\\
&\ \ \ + <X_1X_d G^{d-1}_{k}, X_1 G^{d-1}_{k}>+<X_1X_d G^{d-1}_{k},X_dG^{d-1}_{k+1}>\\
&= 2k + 0 + 0 +  <X_1 G^{d-1}_{k},G^{d-1}_{k+1}>
\end{align*}
Equation (\ref{eq:AlgoCondition}) will thus hold for $2k$ if it holds for $k$. When $(m-1)/2= 2k-1$ we can easily check that the same implication is obtained:
\[
<G_{k-1}^{d-1}, X_1 G_{k}^{d-1}> = 2k-1\Rightarrow
<G_{2k-1}^d, X_1 G_{2k}^d> = 4k-1  \]
Thus, the condition for (\ref{eq:AlgoOddM}) to produce $(d,m)$-edge equitable solutions is 
\begin{equation}
\forall d,\ \forall k <G_{k}^d, X_1 G_{k+1}^d> = 2k+1\enspace .
\end{equation}
It is easy to check that the condition holds for $k=1$ ( $S_d$) and $k=2$ (the composition $(1+X_1X_d)S_d$), which concludes the proof.
\end{itemize}

\section*{Acknowledgment}
  This work has been partially funded by project Desire, ANR, {\em Programme Blanc International II} (FRANCE).

{}


\begin{thebibliography}{}
\bibitem[Boukouvalas (2011)]{bouk:11}
Boukouvalas, A., Gosling, J.P., Maruri-Aguilar, H., (2011).
\newblock An Efficient Screening Method for Computer Experiments.
\newblock  \emph{Tech. Report  NCRG (Aston Univ.)}. 

\bibitem[Campolongo (1999)]{campolongo:99}
Campolongo, F.,  Braddock, R.D.  (1999).
\newblock The use of graph theory in the sensitivity analysis of the model output: a second-order screening method.
\newblock \emph{Reliability Eng. \& System Safety} \textbf{64}, 1--12.

\bibitem[Campolongo(2007)]{campo:07}
Campolongo, F.,  Cariboni, J., Saltelli, A. (2007).
\newblock An effective screening design for sensitivity analysis of large models.
\newblock \emph{Environmental Modelling \& Software} \textbf{22}, 1509--1518.

\bibitem[Campolongo(2011)]{campo:11}
Campolongo, F., Saltelli, A., Cariboni, J. (2011).
\newblock From screening to quantitative sensitivity analysis. A unified approach .
\newblock \emph{Computer Physics Communications} \textbf{182}, 978--988.

\bibitem[Cropp (2002)]{cropp:02}
Cropp, R.A., Braddock, R.D.,  (2002).
\newblock The New Morris method: an efficient second-order screening method.
\newblock \emph{Reliability Eng. \& System Safety} \textbf{78}, 77--83.


\bibitem[Cukier {\em et al} (1973)]{Cukier:73}
Cukier, R.I.,  Fortuin, C.M.,  Shuler, K.E., Petschek, A.G., Schailby, H.,  (1973),
\newblock Study of the 
Sensitivity of the Coupled Reaction Systems to Uncertainties in Rate Coefficients: I. 
Theory.
\newblock  \emph{Journal of Chemical Physics},  \textbf{59}-8, 3873-3878.




\bibitem[Cukier {\em et al} (1978)]{Cukier:78}
Cukier, R.I., Levine, H.B., Schuler, K.E.,  (1978),
\newblock Study of the Nonlinear sensitivity analysis of multiparameter model systems.
\newblock  \emph{Journal of Chemical Physics},  \textbf{26}, 1-42.

\bibitem[Harary (1988)]{Harary:88}
Harary, F., Hayes, J. P., Wu, H.-J(1988).
\newblock A survey of subgraphs of the hypercube.
\newblock  \emph{Comput. Math. Applic.} \textbf{15}-4, 277-289.


\bibitem[Hornberger (1981)]{Horneberger:81}
Horneberger, G.M., Spear, R.C., (1981).
\newblock An approach to the preliminary analysis of environmental systems.
\newblock  \emph{J. Env. Management},  \textbf{12}-7, 7-18.

\bibitem[Morris (1991)]{morris:91}
Morris, M., (1991).
\newblock Factorial Sampling Plans for Preliminary Computational Experiments.
\newblock \emph{Technometrics} \textbf{33}, 161--174.


\bibitem[Pistone (1996)]{Pistone:96}
Pistone, G., Wynn, H., (1996).
\newblock Generalised confiding withGro$\beta$ner Bases.
\newblock  \emph{Biometrika} \textbf{83}-3, 653--666.

\bibitem[Pujol (2007)]{pujol:07}
Pujol, (2007).
\newblock Simplex-based screening designs for estimating metamodels.
\newblock  \emph{Reliability Engineering and System Safety}, \textbf{94}-7, 1156-1160.


\bibitem[Saltelli {\em et al} (2006)]{Saltelli:06}
Saltelli, A., Ratto, M., Tarantola, S., Campolongo, F., (2006).
\newblock Sensitivity Analysis Practices: Strategies for Model Based Inference.
\newblock \emph{Rel. Eng., \& Syst. Safety } \textbf{91}, 1109--1125.

\bibitem[Skinner (1981)]{Skinner:81}
Skinner, C.J., (1981).
\newblock Estimation of the Variance of a Finite Population for Cluster Samples.
\newblock  \emph{Sankhy\={a}: The Indian Journal of Statistics, Series B},  \textbf{43}-3, 392-398.

\bibitem[Sobol (1993)]{Sobol:93}
 Sobol, I.M., (1993).
\newblock Sensitivity estimates for nonlinear mathematical models.
\newblock  \emph{Mathematical Modelling and
Computational Experiments},  \textbf{1}-4, 407-414.

\bibitem[Williams (1999)]{Williams:99}
Williams, R.L., (1999).
\newblock A Note on Robust Variance Estimation for Clustered-Correlated Data.
\newblock  \emph{Biometrika} \textbf{56}, 645-646.











\end{thebibliography}
\end{document}